\documentclass[12pt,a4paper,oneside]{article}

\usepackage{graphicx}
\usepackage{physics}
\usepackage{url}
\usepackage{subfigure}
\usepackage{makecell}
\usepackage{amssymb}

\usepackage{listings}
\newcommand*\lsin{\lstinline[columns=fixed]}

\usepackage[a4paper,left=3cm,right=2cm,top=2.5cm,bottom=2.5cm]{geometry}

\newcommand{\floor}[1]{\lfloor #1 \rfloor}
\begin{document}

\title{Quantum tomography benchmarking}
\author{Bantysh B. I.\footnote{e-mail: bbantysh60000@gmail.com} \and Chernyavskiy A. Yu. \and Bogdanov Yu. I.}
\date{Valiev Institute of Physics and Technology of Russian Academy of Sciences, Moscow, Russian Federation}

\maketitle

\begin{abstract}
Recent advances in quantum computers and simulators are steadily leading us towards full-scale quantum computing devices. Due to the fact that debugging is necessary to create any computing device, quantum tomography (QT) is a critical milestone on this path. In practice, the choice between different QT methods faces the lack of comparison methodology. Modern research provides a wide range of QT methods, which differ in their application areas, as well as experimental and computational complexity. Testing such methods is also being made under different conditions, and various efficiency measures are being applied. Moreover, many methods have complex programming implementations; thus, comparison becomes extremely difficult.

In this study, we have developed a general methodology for comparing quantum state tomography methods. The methodology is based on an estimate of the resources needed to achieve the required accuracy. We have developed a software library (in MATLAB and Python) that makes it easy to analyze any QT method implementation through a series of numerical experiments. The conditions for such a simulation are set by the number of tests corresponding to real physical experiments. As a validation of the proposed methodology and software, we analyzed and compared a set of QT methods. The analysis revealed some method-specific features and provided estimates of the relative efficiency of the methods.

\end{abstract}

\section{Introduction}\label{sect:introduction}

High-precision initialization, transformation and measurement of quantum states is necessary for the practical usage of quantum computers. Control of more than 50 qubits was shown in recent papers \cite{arute2019,bernien2017}, however, the accuracy achieved in these experiments does not allow solving practical problems faster than on classical computers. Additionally, debug methods, including quantum tomography (QT), need to be improved to achieve better characteristics of quantum computational devices \cite{banaszek2013,dariano2003,lvovsky2009}.

All QT methods are defined by the conducted measurements and approaches to processing the obtained results. Many methods are stated as universal ones, but for maximal efficiency QT need to be adapted for specific tasks. For example, non-adaptive QT methods for density matrices are ineffective for the reconstruction of pure and almost pure states. The infidelity of such reconstruction is $\propto 1/N^{1/2}$ \cite{bogdanov2011_1,pogorelov2017,huszar2012,bagan2006,straupe2016}, where $N$ is the sample size. These methods include projected pseudo-inversion \cite{smolin2012}, standard convex optimization \cite{deburgh2008}, methods based on Cholesky decomposition \cite{banaszek1999}, projective gradient descent \cite{bolduc2017,shang2017_1}, etc. Meanwhile, effective methods must produce the $\propto 1/N$ error level \cite{gill2000,bogdanov2009,bogdanov2011_2}. The proportionality constants can differ by an order of magnitude for different methods, which affects their relative efficiency \cite{bantysh2020}.

In addition to the accuracy of the reconstruction, QT methods have other characteristics. For example, some adaptive methods, which provide very high accuracy for some tasks, require large computational resources to calculate optimal measurements, and also require real-time tuning of the measuring devices due to changes in the measurement basis. The variety of QT methods and the lack of comparison methodology make it difficult to build a general picture of the relative efficiency of the methods. Such comparison is also important for the development of new QT methods. However, the task presents some difficulties. Papers on QT consider different QT problems and test conditions, and use different efficiency measures. For example, in the study \cite{flammia2012} on compressed sensing, while the main emphasis in the results is on the comparison of measurement protocols of different dimensions, less attention is paid to the dependence on the sample size. Work \cite{shang2017_1} presents a fast maximal likelihood estimation algorithm for density matrices, but details related to measurement protocols and reconstruction accuracy are omitted. In the paper \cite{bogdanov2009} on the root approach in QT, the method is demonstrated only on a single example of a mixed state. The comparison of adaptive tomography methods in \cite{struchalin2018} is made only for high-dimensional random bipartite states, so the relative efficiency for multipartite states is out of the scope.

This problem is further complicated by the fact that most QT methods are difficult to implement. This does not allow carrying out quick comparative tests for choosing the most efficient method for a specific practical problem.

Note that in other areas of science and technology, the same issues are often solved by unified benchmarks. For example, a wide range of datasets is available for estimating the efficiency of machine learning methods \cite{machlearn}. The analysis of the quality of random number generators is also performed using standardized tests \cite{rukhin2010}. Following this paradigm, in this paper, we propose a set of reference criteria for assessing the quality of QT methods from the point of view of applied problems.

Usually, the analysis and comparison of QT methods is based on the fidelity $F$ that the methods can provide by processing the results of $N$ independent measurements (size of the statistical ensemble). This approach faces the difficulty of choosing a benchmark value $N_B$, since different methods achieve optimal performance at different values of $N$. Instead, we propose the inverse problem: What resources does the QT method need to achieve the benchmark fidelity $F_B$? The proposed methodology is described in Section~\ref{sect:benchmarks}. Based on this methodology, we subject QT methods to a series of tests differing in the type of quantum states being reconstructed. The tests are described in Section~\ref{sect:tests}. For the systematic analysis of QT methods, we have developed easy-to-use software for the MATLAB and Python programming languages (Section~\ref{sect:software}). To demonstrate the capabilities of our approach, we have implemented a fairly wide range of QT methods. Some implementations are based on open software libraries provided by the respective authors. A brief description of the considered methods is presented in Section~\ref{sect:methods}. The results of the analysis and comparison are presented in Section~\ref{sect:examples}. Section~\ref{sect:discussion} provides a discussion of the results obtained. Note, however, that our software implementation of QT methods is based on their descriptions in the respective publications. Adapting these methods to the general tests discussed can improve performance.

\section{Quantum tomography benchmarks}\label{sect:benchmarks}

\subsection{Resources}

The proposed methodology for the analysis and comparison of QT methods is based on an estimate of the resources required to achieve a fixed accuracy of the quantum state reconstruction.

We consider the following resources:
\begin{itemize}
  \item $N$ -- sample size;
  \item $M$ -- number of different measurement bases;
  \item $T_P$ -- time of computing the measurement protocols;
  \item $T_E$ -- time of the reconstruction of a density matrix based on all the measurement results.
\end{itemize}
As a measure of correspondence between the true density matrix $\rho$ and the reconstructed one $\sigma$, we use Uhlmann's fidelity \cite{uhlmann200}:
\begin{equation}\label{eq:fidelity}
  F = \qty(\Tr\sqrt{\sqrt\rho \sigma \sqrt\rho})^2.
\end{equation}
This measure is one of the most commonly used in theory and experiment. It has a number of important properties for mixed states, and in the case when $\rho$ and $\sigma$ describe pure states, the formula \eqref{eq:fidelity} yields the square of the scalar product of the corresponding state vectors \cite{josza1994,liang2019}.

Generally, in each experiment, the QT fidelity is a random variable. Therefore, to analyze QT methods, we carry out $N_e$ runs of a numerical experiment for each value of $N$ from a predefined set. In each experiment, a new random state from the class defined by the considered test is generated (the set of tests is described in Section~\ref{sect:tests}). Then the Monte Carlo measurements are simulated and the density matrix $\sigma$ is reconstructed. The algorithms for selecting a set of measurements and reconstructing a state are determined by the analyzed QT method.

\begin{figure}
  \hfill
  \subfigure[]{\includegraphics[width=.49\linewidth]{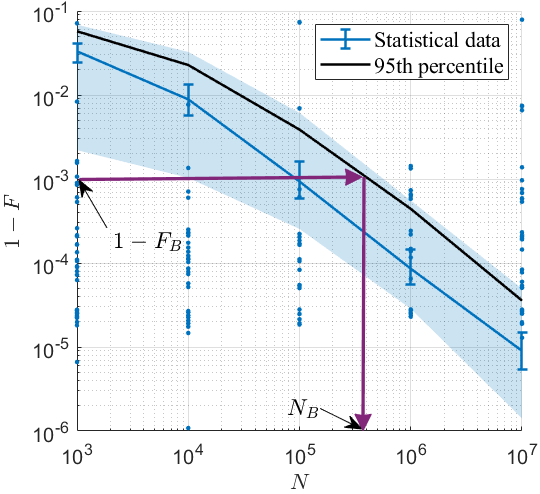}}
  \hfill
  \subfigure[]{\includegraphics[width=.49\linewidth]{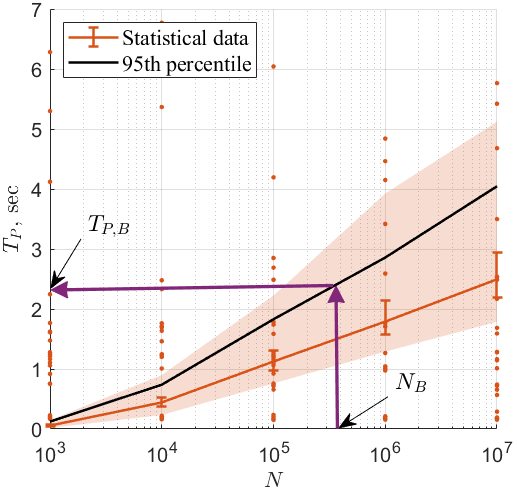}}
  \hfill
  \caption{The method of estimating the reference parameters (benchmarks) using statistical data. (a) Computing the 95th percentile for every value of $N$ (sample size) over $N_e$ runs of a numerical experiment. The benchmark sample size $N_B$ is computed by linear interpolation for the chosen benchmark fidelity $F_B$. (b) Other parameters' benchmarks are computer by their 95-th percentile values in $N_B$.}
  \label{fig:method}
\end{figure}

After the numerical experiments, we calculate the resources that are required for the method to achieve fidelity $F_B$ according to the following algorithm (Fig.~\ref{fig:method}):
\begin{enumerate}
  \item For every numerical experiment defined by $N$ (sample size) and $i$ (run number) compute $F^i(N)$, $M^i(N)$, $T_P^i(N)$ and $T_E^i(N)$.
  \item For every $N$, calculate the 95th percentiles over $i$: $[1-F]_{95}(N)$, $M_{95}(N)$, $T_{P,95}(N)$ and $T_{E,95}(N)$.
  \item Perform linear interpolation of the dependency from the previous Step and find the value $N_B$ corresponding to $[1-F]_{95}(N_B) = 1-F_B$.
  \item Find $M_{95}^B=M_{95}(N_B)$, $T_{P,95}^B=T_{P,95}(N_B)$, $T_{E,95}^B=T_{E,95}(N_B)$ using linear interpolations of dependencies from Step 2.
\end{enumerate}

The computed values of $N_B$, $M_{95}^B$, $T_{P,95}^B$ and $T_{E,95}^B$ are estimates of the resources necessary to achieve a fidelity of at least $F_B$ with a probability of 95\%. Linear interpolation is performed on a logarithmic scale due to the fact that $1-F \propto 1/N^q$, where $0<q\leq1$ \cite{bogdanov2011_1,pogorelov2017,huszar2012,bagan2006,straupe2016,gill2000,bogdanov2009,bogdanov2011_2}. The 95\% probability value has been chosen based on our collaboration with different experimental groups. However, this number can be easily adapted for specific practical tasks.

Note that the times $T_{P,95}^B$ and $T_{E,95}^B$ depend on the computational device used for the analysis. These characteristics can also differ for different algorithmic and programming implementations of the same QT method.

\subsection{Lower bound and efficiency}

Let $d$ be the dimension of the Hilbert space, and $r$ be the rank of the quantum state density matrix. It is known that for asymptotically unbiased estimates of a quantum state obtained by a set of POVM (positive-operator valued measure) measurements and asymptotically efficient methods in terms of the Cram{\'e}r--Rao bound (such as the maximum likelihood method), the QT infidelity has a generalized chi-squared distribution \cite{bogdanov2009,bogdanov2011_2,bogdanov2010}:
\begin{equation}\label{eq:infid_distr}
  1-F \sim \sum_j^\nu{d_j \xi_j^2},
\end{equation}
where $\nu=(2d-r)r-1$ is the number of independent parameters of a quantum state (number of degrees of freedom), $d_j$ are the variances of the estimates of these parameters, $\xi_j$ are independent random variables with a standard normal distribution. The highest fidelity is achieved when $d_1 = \dots = d_\nu = d_0$ \cite{bogdanov2011_2}. Then, for POVM measurements, $d_0=\nu/[4N(d-1)]$ and \eqref{eq:infid_distr} is proportional to the chi-squared distributed random variable with $\nu$ degrees of freedom: ${1-F \sim d_0\chi_\nu^2}$. This allows one to compute the lower bound of the 95th percentile of infidelity by the inverse cumulative distribution function (ICDF) $\mathcal{F}^{-1}$: ${[1-F]_{95}=d_0\mathcal{F}_{\chi^2}^{-1}[0.95|\nu]}$, and also find the corresponding benchmark sample size $N_B$.

It is also useful to introduce the concept of the efficiency of the method in relation to the average infidelity:
\begin{equation}\label{eq:efficiency}
  \eta = \frac{\expval{1-F}_{min}}{\expval{1-F}} = \frac{\nu d_0}{\expval{1-F}} = \frac{\nu^2}{4N(d-1)\expval{1-F}}.
\end{equation}
Efficiency takes the values from 0 to 1\footnote{Some measurements protocols could consider only a part of the measurement events and do not form complete POVM measurements. In this case, assuming $N$ to be the total number of observed events could result in $\eta > 1$ according to the equation \eqref{eq:efficiency}.}. The $\eta_B$ value for $N_B$ is computed by linear interpolation of $\eta(N)$.

\subsection{Outliers}\label{eq:outliers}

The robustness of a QT method with respect to statistical fluctuations can be characterized by the number of outliers in the distribution of infidelity. The standard method for analyzing outliers is based on the quartiles $Q_1$ and $Q_3$ (all points outside the interval $[Q_1-1.5IQR, Q_3+1.5IQR]$ are considered outliers, where $IQR = Q_3 - Q_1$ is the interquartile range). We mentioned above that for efficient QT methods, the infidelity has a generalized chi-squared distribution on condition that for each degree of freedom there is a sufficiently large number of observations. This distribution is significantly asymmetric, and the standard method mentioned above cannot be applied. In this regard, we use a modified approach to determining outliers proposed in \cite{hubert2008}. It is based on the calculation of medcouple ($MC$), which characterizes the slope of a distribution \cite{brys2003}:
\begin{equation}
  MC = \underset{\scriptstyle [1-F]_-<Q_2<[1-F]_+}{\textrm{med}}\qty[\frac{([1-F]_+-Q_2)-(Q_2-[1-F]_-)}{[1-F]_+ - [1-F]_-}].
\end{equation}
Here, $Q_2$ is the sample median of infidelity, $[1-F]_+$ ($[1-F]_-$) are all values of infidelity higher (lower) than $Q_2$, $\textrm{med}[\dots]$ is the median of all values inside the brackets. For a positively skewed distribution of infidelity, we have $MC > 0$. Then all $1-F$ values outside ${[Q_1 - 1.5e^{-4MC}IQR, Q_3 + 1.5e^{3MC}IQR]}$ are considered outliers.

We compute the outliers ratio $O$ (the number of outliers divided by the number of experiment runs) for every $N$, and again $O_B$ corresponding to $F_B$ is computed using linear interpolation of $O(N)$.

\subsection{Additional properties}

QT methods also have some other important properties that can determine their efficiency within the framework of the reference criteria introduced above. An important characteristic is the possibility of implementing the method using only factorized measurements, which measure each subsystem independently of the others. Such measurements are much easier to perform experimentally than entangled measurements.

\section{Benchmark tests}\label{sect:tests}

As part of the developed approach, we propose analyzing QT methods using a set of reference tests (benchmarks). When developing these tests, we tried to cover a sufficiently wide range of experimental situations to track differences in tomography methods efficiency. Further development of the test suite, however, is still an important task (see discussion section).

To accumulate statistics, $N_e = 1000$ numerical experiment runs are performed in each test for each sample size $N$. The values of $N$ are determined by integer powers of 10 (the minimum and maximum values were selected empirically). In each numerical experiment, a random $n$-qubit density matrix of a certain class is generated. Below, the following notation will be used: $\textrm{Haar}(d)$ is a random (according to the Haar measure) unitary matrix of dimension $d \times d$, $\textrm{norm}(0,\sigma)$ is a normally distributed random variable with zero mean and standard deviation $\sigma$, $\textrm{unif}(a,b)$ is a random variable distributed uniformly from $a$ to $b$.

\subsection{Random pure states (RPS)}\label{sect:rps}

In some physical implementations of quantum systems, quantum states of a very high degree of purity ($\Tr(\rho^2) \approx 1$) can be prepared. For relatively small sample sizes, such states can generally be considered pure, since measurements in this case are insensitive to the presence of weak impurities \cite{bogdanov2011_1,struchalin2018}. The rank of the density matrix corresponding to a pure state is equal to unity, and the methods that consider general density matrices then turn out to be inefficient. The RPS test considers the following states:
\begin{equation}
  \rho = \ketbra{\psi}, \quad \ket\psi = U\ket0, \quad U \sim \textrm{Haar}(d).
\end{equation}

\subsection{Random mixed states by partial tracing (RMSPT)}\label{sect:rmspt}

Debugging quantum computations is one of the most important applications of QT. It is often possible to encounter a situation when experimental and computational resources allow one to perform QT of only a separate subsystem $S$ of the composite system $S+A$. This situation may be due to both the complexity of carrying out a sufficient number of measurements over the whole composite system and the complexity of reconstructing the state of a high-dimensional system.

The density matrix of a subsystem $S$ is a partial trace over the subsystem $A$ \cite{nielsen2000}. In this regard, we consider quantum states of the following form for the RMSPT test:
\begin{equation}
  \rho = \Tr_A(\ketbra{\psi}), \quad \ket\psi = U\ket0, \quad U \sim \textrm{Haar}(d_S \cdot d_A),
\end{equation}
where $d_S = d$ and $d_A$ are the dimensions of the subsystems $S$ and $A$, respectively. We will consider two cases: in the first one, the subsystem $A$ consists of a single qubit, so $d_A = 2$ (RMSPT-2), and in the second one, the dimension of $A$ is equal to the dimension of $S$, so $d_A = d_S = d$ (RMSPT-d). Note that the second case is equal to the uniform distribution of density matrices induced by the Hilbert--Schmidt distance \cite{zyczkowski2001}.

\subsection{Random noisy preparation (RNP)}\label{sect:rnp}

This test is aimed at analyzing quantum states that occur often in experiments. Consider the preparation of a quantum state $\ket\psi$ made by the initialization of the $n$-qubit register ${\ket0 = \ket{0}_1 \otimes \dots \otimes \ket{0}_n}$ and applying some unitary transform $U$: $\ket\psi = U\ket0$. In physical systems, the initialization may contain an error $e_0$ (the probability that the $j$th qubit was initialized in the state $\ket{1}_j$). This situation leads to the density matrix of the $j$th qubit: ${\rho_{0,j}^{e_0} = (1-e_0)\ketbra{0}_j + e_0\ketbra{1}_j}$. Then the full register is ${\rho_0^{e_0} = \rho_{0,1}^{e_0} \otimes \dots \otimes \rho_{0,n}^{e_0}}$. The ensuing unitary transform yields ${\rho^{e_0} = U\rho_0^{e_0}U^\dagger}$. If the error is small, then the principal component $\ket\psi$ of $\rho^{e_0}$ corresponds to the eigenvalue $(1-e_0)^n$ that defines the probability of the correct initialization.

Now, let the unitary transform be noisy too. Instead of the ideal $U$ we consider the transform $V = W \cdot U$, where $W$ is a random unitary matrix such that
\begin{equation}\label{eq:randunitary}
  W\ket\varphi = a\ket\varphi + \sqrt{1-a^2}\frac{\ket{g} - \ket{\varphi}\braket{\varphi}{g}}{\norm{\ket{g} - \ket{\varphi}\braket{\varphi}{g}}},
\end{equation}
where $\ket{g}$ is a random vector generated by the Haar measure, $a=1-\xi^2/2$, $\xi \sim \textrm{norm}(0,\sigma)$, and the standard deviation $\sigma$ is a small parameter that characterizes the level of error \cite{struchalin2016,palmieri2020}. It can be shown (see Appendix) that averaging over \eqref{eq:randunitary} is equivalent to adding a depolarizing noise:
\begin{equation}\label{eq:randunitary_depol}
  \int{W \rho W^\dagger dW} = (1-p)\rho + pI_d/d,
\end{equation}
where $p=\frac{d}{d-1}(\sigma^2 - \frac34 \sigma^4)$, $I_d$ is the identity matrix of size $d \times d$. Note that the depolarizing noise is also a good approximation of other types of errors \cite{dur2005}.

In the RNP test, we consider the tomography of random states generated by the procedure described above. The corresponding density matrix is
\begin{equation}
  \rho = (1-p)U\rho_0^{e_0}U^\dagger + pI_d/d, U \sim \textrm{Haar}(d), p \sim \textrm{unif}(0,0.01), e_0 \sim \textrm{unif}(0,0.05).
\end{equation}
We have chosen the parameters $p$ and $e_0$ in the form of uniformly distributed random variables due to the fact that the preparation errors can differ significantly for various physical platforms of quantum computing developed in recent years \cite{arute2019,bernien2017,watson2018,tosi2017,wu2019,wright2019}. Limits $0.01$ and $0.05$ roughly describe the worst performance at the moment.

\section{Analysis software}\label{sect:software}

For the unified analysis of different QT methods, we developed easy-to-use open-source software libraries for the MATLAB \cite{bantysh2020mqtb} and Python \cite{bantysh2020pyqtb} programming languages.

A QT method must be implemented by two \textit{handlers}:
\begin{itemize}
  \item the \textit{measurement protocol handler} taking as input the sequential number of the current copy of the state being reconstructed and returning a description of the measurements to be performed on the following states;
  \item the \textit{estimator handler} taking the results of all previous measurements and returning a reconstructed density matrix. 
\end{itemize}
The described handlers are arguments of the \textit{qtb\_analyze} function. Other arguments are the dimension of the system and the identifier of a test. Generation of states and Monte Carlo simulation of measurements are carried out inside the \textit{qtb\_analyze}. For each numerical experiment, the software computes the fidelity, the number of measurements and other benchmarks described in Section~\ref{sect:benchmarks}. It is important that the handlers do not have access to the true state being reconstructed. The described approach made it possible to make the QT simulation procedure as close as possible to the way QT works in a real experiment.

Listing~\ref{lst:qtb} demonstrates how a two-qubit state QT method is analyzed by the RPS test using just 3 lines of code, and the results are saved to an excel file as a table.

\begin{lstlisting}[caption={An example of calling a function that analyzes a two-qubit QT method in \mbox{MATLAB}. The first line defines the dimensions of each state subsystem. The analysis of the QT method (the FMUB-PPI method) by the RPS test is conducted on the second line. The third line prepares the analysis results according to the benchmarks. The results are dumped as a table into the report.xlsx file},label={lst:qtb},language=Matlab]
dim = [2,2];
result = qtb_analyze(proto_fmub(dim), est_ppi(), dim, 'rps');
qtb_report(result, 'rps', 'export', 'report.xlsx');
\end{lstlisting}

The first line defines the dimension of each subsystem. Note that the case of two qubits differs from the case of a single ququart (when \lsin{dim = [4]}). Despite the fact that in both cases the total dimension of the system is the same, the particular methods implementations could vary.

The second line of the code performs the statistical data collection through numerous implementations of numerical tomography experiments. Each such experiment consists of the following steps:
\begin{enumerate}
\item Generate a quantum state $\rho$ according to the selected benchmark test (Section~\ref{sect:tests}).
\item Request a set of measurements to be performed over $N$ quantum state samples from the measurement protocol handler (\lsin{proto_fmub(dim)} in Listing~\ref{lst:qtb}).
\item Calculate the probabilities of observing the $k$-th result according to the Born rule: $p_k^j=\textrm{Tr}(\rho P_k^j)$ using the POVM operators matrices $P_k^j$ in the $j$-th measuremen.
\item For the $j$-th measurement use the probabilities $p_k^j$ to simulate random outcomes using multinomial statistics.
\item Pass the simulated data and the measurement protocol as the input of estimator handler (\lsin{est_ppi()} in Listing~\ref{lst:qtb}) and obtain the reconstructed density matrix $\sigma$.
\item Calculate the fidelity $F$, the total number of measurement bases $M$, and the protocol and estimator computation times $T_P$ and $T_E$. Write these values into the output data array.
\end{enumerate}
Note that throughout the program the true state $\rho$ is unavailable to the measurement protocol and estimator handlers. The second line of Listing~\ref{lst:qtb}, which was described above, is the most time-consuming. Its duration mainly depends on the speed of the tomography method itself.

Finally, the third line of  Listing~\ref{lst:qtb} takes the obtained statistical data as input, calculates the benchmark values (see Section~\ref{sect:benchmarks}) and exports them in the form of an Excel table.

\section{Methods summary}\label{sect:methods}

Each tomography method differs in the measurement protocol (Section~\ref{sect:protos}) and the way the results of these measurements are processed (Section~\ref{sect:estimators}). At the same time, we will single out three types of measurement protocols most common in the literature:
\begin{itemize}
  \item \textit{POVM} -- each measurement is described by a set of POVM operators \cite{nielsen2000};
  \item \textit{observable} -- measurements of average values of observables;
  \item \textit{operator} -- only a single POVM operator is considered during a single measurement.
\end{itemize}
Since the POVM-type measurements are the most general ones among them, the observable-type and operator-type measurements can be treated as special cases of POVM-type ones.

When specifying a tomography method with a protocol--estimator combination, it is necessary that an estimate can be made based on the considered type of measurements. For example, the MUB-FRML method contains POVM-type measurements in mutually unbiased bases and a maximum likelihood estimation of the full-rank density matrix. The adaptive AMUB-FRML method uses an adaptive MUB measurement protocol, and the density matrix is estimated at each adaptive step using the maximum likelihood method.

Some tomography methods imply only a certain combination of the measurement protocol and density matrix estimation. We will describe them in Section~\ref{sect:other_methods}. 

In addition to the methods below, there are approaches that may be of interest, but have not yet been implemented within the software interface: Bayesian inference tomography \cite{huszar2012}, adaptive compressed sensing tomography \cite{ahn2019}, pure states tomography using five bases protocols \cite{goyeneche2015}, machine learning methods \cite{torlai2018,fastovets2019,palmieri2020}, continuous-variable tomography \cite{lvovsky2009,bogdanov2016} and some others.

Note that here we do not focus on the numerical algorithms used in the implementation of methods, since they practically do not affect the main benchmarks of the method except the times $T_P$ and $T_E$.

\subsection{Measurements protocols}\label{sect:protos}

\subsubsection{Mutually unbiased bases (MUB)}\label{sect:mub}

Protocols with some symmetry are of particular interest for tomography of unknown quantum states. These protocols include measurements in mutually unbiased bases \cite{bengtsson2007,adamson2010}. Let state vectors $\ket{e_k}$ and $\ket{f_k}$ ($k = 0, \dots, d-1$) form two orthonormal bases. They are called mutually unbiased if the following relations are satisfied:
\begin{equation}\label{eq:mub}
  \abs{\braket{e_j}{f_k}}^2 = \frac{1}{d}, \quad j,k = 0, \dots, d-1.
\end{equation}
Here, by a MUB protocol we mean a set of the maximum number of bases satisfying \eqref{eq:mub}. In this case, it is assumed that the vectors of one of the bases form the identity matrix. The maximum number of MUBs is $d+1$ for a space whose dimension is a power of a prime. This protocol is of POVM-type. Note that for a qubit ($d=2$) any MUB protocol is equivalent to projective measurements in three bases corresponding to the Pauli operators up to unitary rotation.

\subsubsection{Factorized mutually unbiased bases measurements (FMUB)}\label{sect:fmub}

Factorized measurements of subsystems are of particular interest, since they are most easily realized experimentally. By a factorized MUB protocol we mean a POVM-type measurement protocol in which each subsystem is measured independently using the MUB protocol (Section~\ref{sect:mub}). The number of bases in such a protocol is $\prod_j{(d_j+1)}$, where $d_j$ is the dimension of the $j$th subsystem.

\subsubsection{Pauli measurements (Pauli)}\label{sect:pauli}

This observable-type protocol is based on the independent measurement of single-qubit observables corresponding to Pauli operators $\sigma_x$, $\sigma_y$ and $\sigma_z$, complemented by an identity operator $\sigma_0$. In the case of $n$ qubits, such a protocol is formed by all possible tensor products of these operators: $\{\sigma_0, \sigma_x, \sigma_y, \sigma_z\}^{\otimes n} / \sigma_0^{\otimes n}$. Operator $\sigma_0^{\otimes n}$ is excluded because it does not provide any useful information about a quantum state. So, the considered protocol contains $4^n-1$ measurements. The dimension of the state space is $d = 2^n$.

Note that for tomography of rank-deficient states, the number of measurements can be reduced \cite{flammia2012}, however, here we will consider a complete set of $4^n-1$ measurements, since the systems under study have a relatively low dimension.

\subsubsection{Adaptive mutually unbiased bases measurements (AMUB)}\label{sect:amub}

It was shown in \cite{struchalin2018} that, in the case of almost pure states, measurements in the eigenbasis of the state are optimal. Since the eigenbasis is initially unknown, the measurement basis should be selected adaptively: at each step, the quantum state is estimated and its eigenbasis is determined. To ensure the information completeness, the eigenbasis is complemented by other bases, which together form MUB (Section~\ref{sect:mub}).

\subsubsection{Factorized orthogonal measurements (FO)}\label{sect:fo}

The AMUB protocol generally requires measurements in the entangled states bases. In \cite{struchalin2018}, it was proposed that a factorized non-entangled basis can be constructed containing a vector orthogonal to $K$ principal components of the current estimate of the density matrix. At each adaptive step, the number $K$ is chosen randomly from $1$ to $K_{max}=\sum_{j=1}^{n}{d_j}-n$, where $d_j$ is the dimension of the $j$th subsystem and $n$ is the number of subsystems. The vector resulting from minimization is randomly complemented to a factorizable complete basis. Following \cite{struchalin2018}, in this study, $\max(100, \floor{N_t/30})$ measurements are performed at each adaptive step, where $N_t$ is the total number of representatives measured in the previous steps.

Note that this method cannot be applied to the tomography of a single qubit, since the complement to the full basis will always be unambiguous, and the protocol formed by iterations will not be able to provide informational completeness.

\subsubsection{Factorized orthogonal mutually unbiased bases measurements (FOMUB)}

This protocol, proposed in \cite{bantysh2020}, is a combination of the FO (Section~\ref{sect:fo}) and FMUB (Section~\ref{sect:fmub}) protocols. As in the FO protocol, at each adaptive step, a factorized vector $\ket{\varphi}_1 \otimes \dots \otimes \ket{\varphi}_n$ orthogonal to the $K$ principal components of the current density matrix estimate is being computed. Further, an FMUB protocol is formed such that one of the vectors of one of the bases corresponds to $\ket{\varphi}_j$ for each $j$th subsystem. Thus, one vector in a certain basis of the FMUB turns out to be orthogonal to $K$ principal components of the current density matrix estimate, and the other bases provide informational completeness. However, unlike in the AMUB protocol (Section~\ref{sect:amub}), all the measurements remain factorizable.

\subsection{Density matrix estimators}\label{sect:estimators}

\subsubsection{Projected pseudo-inversion estimator (PPI)}

The simplest and fastest method to implement is the method of pseudo-inversion of frequencies of events obtained in the experiment \cite{bogdanov2011_2}. Due to statistical fluctuations, such a method often gives a matrix $X$ that is not positive semidefinite. To obtain a density matrix, the vector of eigenvalues of the matrix $X$ is projected onto a standard simplex (non-negative numbers with unit sum) \cite{chen2011}. This approach is considered, for example, in \cite{smolin2012}. For such projection, we use a subroutine of an open software library for the MATLAB language for the reconstruction of the density matrix using projective gradient descent \cite{shang2017_2}.

\subsubsection{Full rank least squares estimator (FRLS)}\label{sect:frls}

Reconstruction of an arbitrary full-rank density matrix can be performed by the minimization of the sum of squares between the observed and theoretical frequencies of events under the constraint that the density matrix must be positive semidefinite \cite{kosut2004}. Such a task can be accomplished by convex optimization, for the implementation of which we use the open software library CVX for the MATLAB language \cite{grant2013}. Note that the least squares method under certain conditions is equivalent to pseudo-inversion \cite{kendall1961}.

\subsubsection{Full rank maximum likelihood estimator (FRML)}

Under certain rather general conditions, the maximum likelihood method has optimal asymptotic properties \cite{kendall1961}. Reconstruction of a full-rank density matrix by this method can be performed using a number of equivalent algorithms. Among them are algorithms based on the root approach (for a full rank) \cite{bogdanov2009,bogdanov2010}, Cholesky decomposition \cite{banaszek1999} and projective gradient descent \cite{bolduc2017,shang2017_1}. To reconstruct the density matrix using the root approach, we used open-source software libraries for MATLAB \cite{bantysh2019mrt} and Python \cite{bantysh2019pyrt}.

\subsubsection{True rank maximum likelihood estimator (TRML)}

In real experiments, rank-deficient quantum states are often prepared. The density matrix of such a state contains one or more zero eigenvalues. Knowing the rank $r_t$ of the true density matrix, one can optimize the reconstruction of the quantum state by restricting the search space to matrices of rank $r_t$ only. For this, we used the root approach based on open-source software libraries \cite{bantysh2019mrt,bantysh2019pyrt}. This can also be implemented on the basis of projective gradient descent with the projection of the vector of eigenvalues onto a standard simplex of lower dimension \cite{struchalin2018}.

\subsubsection{Adequate rank maximum likelihood estimator (ARML)}

An experiment often lacks \textit{a priori} information about the rank of a state. In this case, the estimation of the true rank $r_e$ can be carried out using the chi-squared test \cite{bogdanov2009}. For this, the values of rank $r$ from $1$ to $d$ were considered in sequence, the state was reconstructed, and the \textit{p-value} $P_r$ of the model was estimated according to the chi-squared test. If for some value of the rank the condition $P_r \geq \alpha$ was satisfied ($\alpha = 5$\% is the significance level) then the procedure was stopped and the rank $r_e = r$ was chosen as the estimate of the true one. The procedure was also stopped if $P_{r+1} < P_r$ (then $r_e = r$) \cite{bantysh2020}.

\subsubsection{Compressed sensing estimator (CS)}

The compressed sensing method is based on the assumption that the density matrix has a sufficient number of small eigenvalues. In this case, the FRLS estimation (Section~\ref{sect:frls}) is supplemented by minimizing the trace of the density matrix, which is equivalent to minimizing the cardinality (number of non-zero elements) of the vector of the density matrix eigenvalues and, accordingly, the density matrix rank \cite{fazel2004}.

For the observable-type Pauli measurement protocol (Section~\ref{sect:pauli}), we use the Lasso method from \cite{flammia2012}. Its software implementation is based on the MATLAB source code provided by the authors in the supplementary materials.

For POVM-type protocols, the algorithm from \cite{steffans2017} is used. The value ${\varepsilon = \hat\varepsilon(1+\alpha)}$ is considered as an optimization parameter, where $\alpha$ increases from $0$ with a step of $0.1$ until convergence.

\subsection{Other methods}\label{sect:other_methods}

\subsubsection{Self-guided quantum tomography (SGQT)}

This operator-type adaptive method is an experimental computation of the gradient of the value characterizing the proximity of the state vector estimate to the true pure state \cite{ferrie2014,chapman2016}. To analyze the algorithm, the following set of optimization parameters was considered: $A = 0$, $a = 3$, $b = 0.1$, $s = 0.602$, $t = 0.101$, $N = 100$.

\section{Examples of methods comparison}\label{sect:examples}

Using the developed software, we analyzed the methods described above using examples of systems of one, two and three qubits. The full analysis results are given in Supplementary material I. Supplementary material II reports a complete comparison of the methods.

Here we demonstrate with an example how our software allows performing a descriptive quantitative comparison of various methods. To do this, we consider the tomography of two-qubit random pure states (RPS test, Section~\ref{sect:rps}). Table~\ref{tab:rps} shows the benchmarks of various methods at $F_B = 99.9$\%: the total sample size $N_B$, the number of measurements $M_{95}^B$, the calculation time of the protocol $T_{P,95}^B$, the time of calculating the density matrix $T_{E,95}^B$, the efficiency $\eta_B$, the outliers ratio $O_B$, and whether the protocol relies on factorized measurements only ($FM$ is ``Y'' for factorized measurements, ``N'' for non-factorized). Similar data is presented in Table~\ref{tab:rnp} for random noisy preparation test (RNP test, Section~\ref{sect:rnp}).

Depending on method, the data presented in Tables~\ref{tab:rps} and \ref{tab:rnp} were obtained on two different computational devices: a personal computer with two cores processor and a 96 processors cluster. For more information, see Supplementary material I.

\begin{table}
  \caption{Comparison of benchmarks of tomography methods for two-qubit pure states (reaching benchmark fidelity $F_B = 99.9$\% with 95\% probability)}
  \label{tab:rps}
  \begin{tabular}{llllllll}
    \hline\noalign{\smallskip}
     & $N_B$ & $M_{95}^B$ & $T_{P,95}^B$, sec & $T_{E,95}^B$, sec & $\eta_B$ & $O_B$ & $FM$ \\
    \noalign{\smallskip}\hline\noalign{\smallskip}
    \textbf{Lower bound} & 6 296 & -- & -- & -- & 1 & -- & -- \\
    \textbf{FMUB-TRML} & 7 205 & 9 & 0.0014 & 0.0081 & 0.9 & 0.0098 & Y \\
    \textbf{FMUB-ARML} & 7 798 & 9 & 0.0014 & 0.013 & 0.57 & 0.022 & Y \\
    \textbf{PAULI-CS} & 20 239 & 15 & 0.0015 & 0.61 & 0.31 & 0.0071 & Y \\
    \textbf{AMUB-FRML*} & 23 472 & 98 & 2.9 & 0.31 & 0.35 & 0.0026 & N \\
    \textbf{FO-FRML*} & 164 734 & 153 & 170 & 0.14 & 0.062 & 0.048 & Y \\
    \textbf{FOMUB-FRML*} & 172 765 & 172 & 15 & 1.1 & 0.073 & 0.0078 & Y \\
    \textbf{SGQT*} & **1 048 989 & -- & -- & -- & -- & -- & N \\
    \textbf{FMUB-CS} & **1 123 593 & -- & -- & -- & -- & -- & Y \\
    \textbf{FMUB-FRML} & **1 424 621 & -- & -- & -- & -- & -- & Y \\
    \textbf{MUB-FRML} & **1 638 550 & -- & -- & -- & -- & -- & N \\
    \textbf{FMUB-FRLS} & **1 763 979 & -- & -- & -- & -- & -- & Y \\
    \textbf{FMUB-PPI} & **1 811 355 & -- & -- & -- & -- & -- & Y \\
    \noalign{\smallskip}\hline\noalign{\smallskip}
    \multicolumn{8}{l}{\makecell[cl]{* Adaptive method \\ ** The value is obtained by linear extrapolation of the dependence of $\log[1-F]_{95}$ on $\log N$}}
  \end{tabular}
\end{table}

\begin{table}
  \caption{Comparison of benchmark parameters of tomography methods for two-qubit states with noisy preparation (reaching benchmark fidelity $F_B = 99.9$\% with 95\% probability)}
  \label{tab:rnp}
  \begin{tabular}{llllllll}
    \hline\noalign{\smallskip}
     & $N_B$ & $M_{95}^B$ & $T_{P,95}^B$, sec & $T_{E,95}^B$, sec & $\eta_B$ & $O_B$ & $FM$ \\
    \noalign{\smallskip}\hline\noalign{\smallskip}
    \textbf{Lower bound} & 31 245 & -- & -- & -- & 1 & -- & -- \\
    \textbf{AMUB-FRML*} & 165 667 & 161 & 5.2 & 0.2 & 0.28 & 0.0083 & N \\
    \textbf{FOMUB-FRML*} & 629 755 & 218 & 36 & 2.1 & 0.095 & 0.036 & Y \\
    \textbf{MUB-FRML} & 1 016 042 & 5 & 0.00067 & 0.022 & 0.072 & 0.056 & N \\
    \textbf{FMUB-FRML} & 1 251 861 & 9 & 0.00051 & 0.18 & 0.06 & 0.039 & Y \\
    \textbf{FMUB-TRML} & 1 251 861 & 9 & 0.00051 & 0.18 & 0.06 & 0.039 & Y \\
    \textbf{FMUB-FRLS} & 1 382 130 & 9 & 0.0011 & 0.66 & 0.05 & 0.038 & Y \\
    \textbf{FMUB-PPI} & 1 400 187 & 9 & 0.00061 & 0.00057 & 0.049 & 0.038 & Y \\
    \textbf{FMUB-CS} & 2 234 214 & 9 & 0.00091 & 0.49 & 0.027 & 0.014 & Y \\
    \textbf{FO-FRML*} & 2 851 554 & 240 & 301 & 3.5 & 0.026 & 0.014 & Y \\
    \textbf{FMUB-ARML} & 6 238 296 & 9 & 0.0012 & 0.47 & 0.017 & 0.13 & Y \\
    \textbf{PAULI-CS} & **39 664 389 & -- & -- & -- & -- & -- & Y \\
    \noalign{\smallskip}\hline\noalign{\smallskip}
    \multicolumn{8}{l}{\makecell[cl]{* Adaptive method \\ ** The value is obtained by linear extrapolation of the dependence of $\log[1-F]_{95}$ on $\log N$}}
  \end{tabular}
\end{table}

\begin{figure}
  \hfill
  \subfigure[]{\includegraphics[width=.49\linewidth]{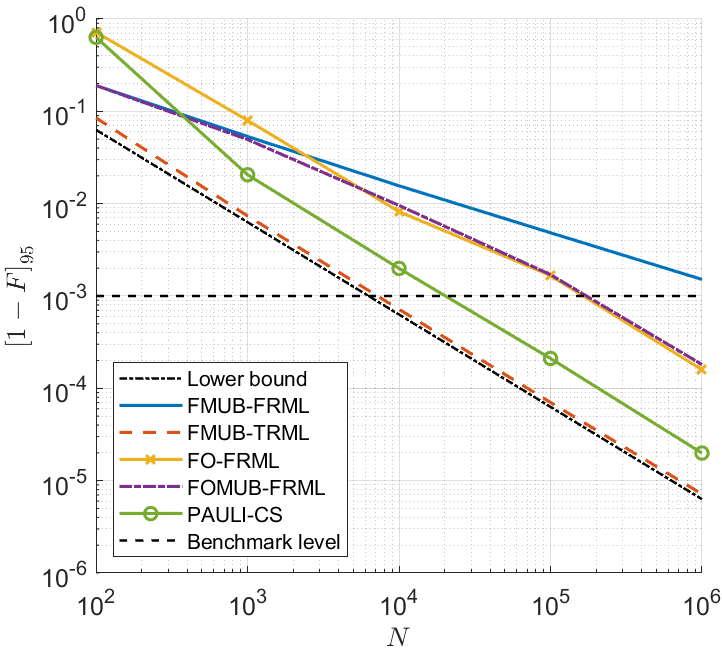}}
  \hfill
  \subfigure[]{\includegraphics[width=.49\linewidth]{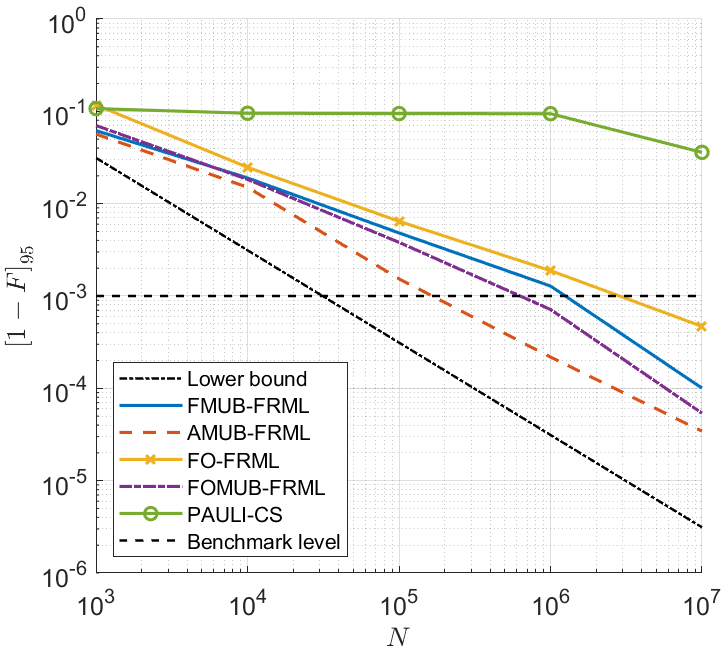}}
  \hfill
  \caption{Dependence of the 95th percentile of the error on the sample size for various methods of tomography of the two-qubit state: RPS test (a) and RNP test (b)}
  \label{fig:results}
\end{figure}

The presented results clearly demonstrate the advantages and disadvantages of various methods in relation to tomography of two important types of states. For example, any methods based on the reconstruction of a general density matrix (without rank restrictions) and using static measurement protocols (MUB-FRML, FMUB-FRML, FMUB-FRLS, FMUB-PPI) require a much larger sample size to ensure the same accuracy in the RPS test than the reduced-rank (FMUB-TRML, FMUB-ARML, PAULI-CS) and adaptive methods (AMUB-FRML, FO-FRML, FOMUB-FRML). Figure~\ref{fig:results}a clearly shows that the infidelity decreases with increasing sample size as $1/N$ for some methods (FMUB-TRML, FO-FRML, FOMUB-FRML, PAULI-CS), and that corresponds to optimal convergence. The proportionality constants, however, are somewhat different. At the same time, for the non-optimal FMUB-FRML method, we have a $1/N^{1/2}$ dependence.

The RNP test is a very difficult case for tomography, when the density matrix contains a large number of small eigenvalues. Methods for reconstructing the full-rank density approach a $1/N$ dependence only with the accumulation of sufficiently large statistics (Figure~\ref{fig:results}b). Using an adaptive tomography strategy can provide faster convergence. However, such methods require a large number of different measurements and substantial computational resources.

\section{Discussion}\label{sect:discussion}

The unified test-based approach to evaluating quantum tomography methods provides a tool for quantitative analysis and comparison of various methods. The advantages of this approach are its versatility and closeness to a real experiment. Any method designed to experimentally perform quantum tomography can be easily analyzed using the developed software.

Perhaps the most significant drawback of the proposed methodology is the high computation time. To analyze the method within the framework of every single test, we have conducted 1000 numerical experiments for each value of the total sample size. When analyzing adaptive methods, each numerical experiment can take tens of seconds. So, the analysis of a three-qubit FO-FRML, when parallelizing numerical experiments on 96 processors, took about three days.

Nevertheless, in our opinion, it is the approach based on a large number of numerical experiments that makes it possible to reveal the real features of the method important for its practical implementation.

At the same time, there are still a number of tasks that are planned to be performed in future versions of the software:
\begin{enumerate}
  \item \textit{Ability of using a priori information about the approximate form of a quantum state.} In real experiments, an approximate form of a quantum state is often known, which can be some pure state that had to be prepared. In this case, the task of tomography is to refine the density matrix of the state, to reveal the structure of the noise, etc.
  \item \textit{Creation of tests with measurement errors.} The tests presented in this paper (see Section~\ref{sect:tests}) consider the measurement procedure as an ideal one. In reality, the presence of systematic measurement errors strongly affects the achievable accuracy of tomography. Some methods, however, are able to effectively take into account and compensate for these errors. Therefore, the introduction of such tests will make it possible to better characterize the methods of quantum tomography in relation to realistic experimental conditions.
  \item \textit{Ability to train tomography models.} Some methods may be based on the preliminary training of the tomography model. In the process of training, not only a model of a quantum state can be built, but also a model of measurement errors, which will allow methods to effectively take into account and suppress them.
  \item \textit{Analysis of tomography methods for quantum processes.} Analysis of transformations performed on quantum systems provides additional opportunities for debugging quantum computations. Many methods of quantum process tomography (QPT) are known. Since the tomography of a process is in many ways similar to the tomography of states, the calculation of the reference parameters of the methods can be performed on the basis of the software we have created.
\end{enumerate}

It is of particular interest to make the tests that fully reflect the practical conditions in which quantum tomography techniques are applied. In this regard, we encourage the scientific community to provide constructive comments to the test formulation.

Solving the above tasks will expand the number of methods that can be analyzed and make the tests closer to real experimental problems. At the same time, we do not expect a significant increase in the computational complexity of the benchmarking procedure since it is the method implementation itself that takes most of the computation time.

Benchmarking QPT might take more computation time since quantum processes are characterized with a much higher number of independent parameters and, therefore, require a more complex procedure. However, from the point of view of quantum computing, only one- and two-qubit quantum gates are usually of interest. The corresponding QPT methods are presented in many papers (e.g. \cite{pogorelov2017,wu2013,knee2018,bantysh2019,huang2019}).

\section{Conclusions}\label{sect:conclusions}

Quantum tomography is one of the key approaches to debugging quantum computations (both, hardware and software parts). There are some very general principles for proving the optimality of a particular method. One of them is to show that the method provides a decrease in infidelity with an increase in the sample size according to the $1/N$ law. In some cases, such convergence may not be achieved immediately, and the proportionality constant may differ significantly for different methods. It is also worth paying attention to other features of the method: the complexity of its experimental implementation, computation time, number of statistical outliers, etc.

All these features can be identified only using numerical simulation. In this work, we have proposed a methodology for the quantitative analysis of quantum tomography methods and developed software that allows estimating the benchmarks of an arbitrary method. The software is based on a system of unified tests that simulate various real-life experimental conditions.

The capabilities of the software were demonstrated using 12 different tomography methods. Their implementation was carried out on the basis of the relevant publications. Note, however, that optimization of the parameters of certain methods may improve some of their characteristics.

\section*{Acknowledgements}
This work was supported by Program of the Ministry of Science and Higher Education of Russia (no. 0066-2019-0005) for Valiev Institute of Physics and Technology of RAS, and by Theoretical Physics and Mathematics Advancement Foundation ``BASIS'' (Grant No. 20-1-1-34-1). We are also grateful to G.I. Struchalin for help in carrying out the computations, to Dr. D.O. Sinitsyn for valuable advice and comments, and to all the experimenters who helped us in developing the set of tests.

\section*{Appendix. Equivalence between random unitary error and depolarization}

Let us show the validity of \eqref{eq:randunitary_depol} for the state $\ket0$ (due to the unitary invariance of the distribution of random states $\ket{g}$, the proof for any state $\ket\varphi=U_\varphi\ket0$ with some unitary operator $U_\varphi$ is carried out in a similar way).

\textit{Theorem.} Let the action of a random unitary operator $W_{g,a}$ on the state $\ket0$ in a Hilbert space of dimension $d$ be given by the expression
\begin{equation*}
  W_{g,a}\ket0 = a\ket0 + \sqrt{1-a^2}\frac{\ket{g}-g_0\ket0}{\sqrt{1-\abs{g_0}^2}},
\end{equation*}
where $a$ is a fixed non-negative parameter and $g_0=\braket{0}{g}$. If vectors $\ket{g}$ are uniformly distributed according to the Haar measure, then the following equality holds
\begin{equation*}
  \int{W_{g,a} \ketbra{0} W_{g,a}^\dagger dg} = (1-p_a)\ketbra{0} + p_aI_d/d,
\end{equation*}
where $p_a = \frac{d}{d-1}(1-a^2)$.

\textit{Proof.} The matrix representation of the vector $W_{g,a}\ket{0}$ in the computational basis has the form $\mqty(a & \sqrt{1-a^2}\mathbf{\tilde{g}}^T)^T$, where the column vector $\mathbf{\tilde{g}}$ specifies the amplitudes of a uniformly distributed random vector (according to the Haar measure) in the space of dimension $d-1$. The corresponding density matrix has the form
\begin{equation*}
  \begin{pmatrix}
    a^2 & \quad & a\sqrt{1-a^2}\mathbf{\tilde{g}}^\dagger \\[0.7em]
    a\sqrt{1-a^2}\mathbf{\tilde{g}} & \quad & (1-a^2)\mathbf{\tilde{g}}\mathbf{\tilde{g}}^\dagger
  \end{pmatrix}.
\end{equation*}
Since the amplitudes of the vector $\mathbf{\tilde{g}}$ are given by normalized complex random variables with standard normal distribution, the expected value of $\mathbf{\tilde{g}}$ is equal to the zero vector. Moreover, the averaging over $\mathbf{\tilde{g}}\mathbf{\tilde{g}}^\dagger$ is proportional to the identity matrix. Thus,
\begin{equation*}
  \int{W_{g,a} \ketbra{0} W_{g,a}^\dagger dg} = \int{\begin{pmatrix}
    a^2 & \quad & a\sqrt{1-a^2}\mathbf{\tilde{g}}^\dagger \\[0.7em]
    a\sqrt{1-a^2}\mathbf{\tilde{g}} & \quad & (1-a^2)\mathbf{\tilde{g}}\mathbf{\tilde{g}}^\dagger
  \end{pmatrix}d\tilde{g}} = \begin{pmatrix}
    a^2 & \quad & \mathbf{0}^T \\[0.6em]
    \mathbf{0} & \quad & \frac{1-a^2}{d-1}\mathbf{I}_{d-1}
  \end{pmatrix}.
\end{equation*}

In our case $a=1-\xi^2/2$, where $\xi \sim \textrm{norm}(0, \sigma)$, and $\sigma$ is a small parameter characterizing the error level ($\sigma$ should be small enough to make $a$ positive almost certainly). Having calculated the expected value ${p = \expval{p_a}_\xi = \frac{d}{d-1}(\sigma^2 - \frac34 \sigma^4)}$, we obtain the equality \eqref{eq:randunitary_depol}.

\end{document}